\documentclass[twocolumn,showpacs,preprintnumbers,amsmath,amssymb]{revtex4}
\usepackage{graphicx}
\usepackage{dcolumn}
\usepackage{bm}
\usepackage{graphicx}
\begin{document}
\title{Condensation Energy and High $T_c$ Superconductivity} 
\author{D. van der Marel$^1$, A. J. Leggett$^2$, J. W. Loram$^3$, J. R. Kirtley$^4$} 
\address{
$^1$Material Science Center, University of Groningen, 9747 AG Groningen, The Netherlands\\
$^2$IRC in Superconductivity, University of Cambridge, Cambridge CB3 0HE, United Kingdom\\
$^3$Department of Physics, University of Illinois at Champaign, Urbana IL 61801-3080\\
$^4$IBM T. J. Watson Research Center, Yorktown Heights, New York 10598}
\date{july 7, 2002}
\begin{abstract} 
From an analysis of the specific heat of one of the cuprate superconductors 
it is shown, that even if a large part of the experimental
specific heat associated with the superconducting phase transition 
is due to fluctuations, this part must be counted when one tries to
extract the condensation energy, $E_{cond}$, from the data. 
Previous work where the fluctuation part was subtracted,
has resulted in an incorrect estimation of $E_{cond}$.
\end{abstract}
\maketitle
In conventional metals superconductivity has been understood to 
result from an effective attractive interaction between electrons. 
This simultaneously causes a reduction of the interaction energy 
and an increase of the kinetic energy when the material becomes 
superconducting. For high $T_c$ superconductors, and in particular 
for the cuprate family, it has been proposed that the opposite 
situation could exist, where the superconducting state is accompagnied 
by a reduction of the charge carrier kinetic energy\cite{cka98,hirsch92,sudip93,pwa95}. 
Experimental investigations of the kinetic energy component perpendicular to the 
superconducting planes of the cuprate high Tc superconductors have 
previously demonstrated that the kinetic energy reduction perpendicular 
to the layers is far too small to account for the condensation 
energy\cite{pwa95,loram94,leggett96}, which ruled out interlayer 
tunneling (ILT) as a mechanism of 
superconductivity\cite{marel96,leggett96,schuetzmann97,moler98,pwa98,tsvetkov98,kirtley98}.
Although evidence was subsequently reported\cite{basov99,basov01} that the c-axis 
kinetic energy is reduced in the superconducting state 
in a number of cases, the amount of energy is orders of magnitude
smaller than earlier estimates of the condensation energy\cite{pwa95,loram94,leggett96}. 

Later experiments have identified two
contributions to the internal energy of cuprate superconductors:
(i) From a re-analyses of inelastic neutron scattering data it
was concluded that the $ab$-plane spin-correlation energy 
was lowered by an amount which may be sufficient to account for 
the condensation energy\cite{zhang}, and 
(ii) an even larger lowering of $ab$-plane kinetic energy was measured 
with optical spectroscopy\cite{molegraaf}.
Since spin-correlations result from exchange interactions,
which in turn reflect the spin-dependent virtual motion of electrons,
these two channels of condensation energy may have at least in part 
the same microscopic origin.

However, the analysis providing the condensation energy from specific
data has been questioned in 1999 by Chakravarty, Kee, and 
Abrahams\cite{cka98} (hereafter CKA), who stated that ``the attempt to extract the 
condensation energy from the specific heat data runs into ambiguity, 
except within a mean field treatment. In the presence of fluctuations, 
superconducting correlations, which can primarily be of in-plane origin, 
contribute to the energy and significantly to the specific heat of 
the normal state.'' In order to resolve this ambiguity, CKA proposed 
to ``subtract the fluctuation effects and to use the remainder as an 
effective specific heat from which to extract the $c$-axis contribution 
to the condensation energy.  The rationale is that free energy can be  
decomposed into a singular and a non-singular part. The universal 
singular part is more sensitive to collective long-wavelength 
fluctuations, while the non-singular part is dominated by short distance 
microscopic pairing correlations.''  Subtracting the fluctuation led
CKA to a value of the condensation energy which was 40 times
smaller than the value obtained in Refs.~\onlinecite{pwa95,loram94,leggett96}.

Here we demonstrate that the analysis of CKA was
internally inconsistent. If carried out correctly,
the subtraction of fluctuation energy makes only a factor
of 2 to 3 difference compared to Refs.~\onlinecite{pwa95,loram94,leggett96}.
Moreover, we show that it is overwhelmingly natural to count also the 
fluctuation contribution in the condensation energy.
 
The condensation energy, $E_{cond}$ is the internal energy of the equilibrium 
phase relative to the internal energy of the normal state. The former is 
the experimentally observed phase, which is superconducting for $T < T_c$, whereas 
the  latter corresponds to the state where all superconducting correlations 
have been suppressed in the  sense that the two-particle correlation function 
tends to zero as a function  of the 'center-of-mass' variable over a range 
no greater than a few times the inter-particle spacing.  
In the remainder of this paper we will use the sub-index $n$ to indicate 
the thermodynamic quantities corresponding to this normal state.  
In any superconductor long range phase-coherence is only present for 
$T < T_c$. In BCS theory long range phase coherence and 
pair-correlations become non-zero simultaneously for $T \le T_c$. 
Knowledge of equilibrium and normal specific heat for $T < T_c$ 
then suffices to determine $E_{cond}$. On the other hand, pair-correlations 
can in principle still exist for temperatures above the transition 
temperature, and indeed such correlations are often associated with 
the pseudo-gap phenomenon in underdoped cuprates. A measurement of
the internal energy, released when the superconducting state is formed
out of the normal state, should now also include the pair-correlations 
which already exist above the superconducting phase transition.
Since our discussion is most easily formulated in terms of
the entropy, let us remind the reader that the entropy follows
uniquely from the specific heat according to the relation
\begin{equation} 
  S(T)=\int_0^T \frac{C(T')}{T'} dT'
 \label{entropy} 
\end{equation} 
{\em If} the temperature dependence of the specific heat is known in 
equilibrium and in the normal state, the free and internal energy 
differences can be calculated directly using the relations 
\begin{eqnarray} 
 F_n(T)-F(T)=\int_T^{\infty}\left(S_n(T')-S(T')\right)dT' \\
 E_n(T)-E(T)=\int_T^{\infty}\left(C(T')-C_n(T')\right)dT' 
 \label{eq:etot+efree}
\end{eqnarray}
The integration limits ensure that $\lim_{T\rightarrow\infty}F(T)=F_n(T)$ 
and $\lim_{T\rightarrow\infty}E(T)=E_n(T)$, in accordance with the
assumption that for $T\rightarrow\infty$ all superconducting 
correlations vanish. The condensation energy corresponds 
to the zero-temperature energy difference $E_{cond}=E_n(0)-E(0)=F_n(0)-F(0)$. 
This positive energy can be obtained either by integrating the 
specific heat difference,
\begin{equation}
 E_{cond}=\int_0^{\infty}\left(C(T)-C_n(T)\right)dT 
 \label{eq:econdF}
\end{equation}
or by integrating the entropy difference 
\begin{equation}
 E_{cond}=\int_0^{\infty}\left(S_n(T)-S(T)\right)dT 
 \label{eq:econdE}
\end{equation}
The fact that Eq. \ref{eq:econdF} and Eq. \ref{eq:econdE} should
give the same value provides, as we will see, an important 
consistency check.
\begin{figure} 
        \centerline{\includegraphics[width=6cm,clip=true]{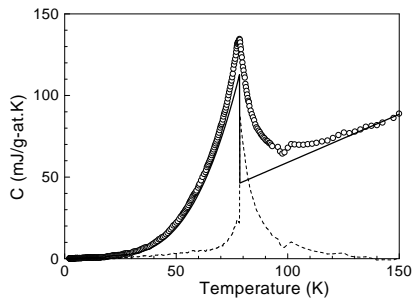}} 
        \caption{Open circles: Electronic specific heat data of Tl2201.
         Dashed curve: Singular (fluctuation) contribution, parametrized as
         $C_{sing}={g_{\pm}/t}$, where $t=|1-T/T_c|$ ($T_c =$ 79 K$, g_+=2.38$, $g_-=0.74$).
        Solid curve: Electronic specific heat with the singular contribution subtracted. } 
\label{fig:C(T)} 
\end{figure} 
CKA\cite{cka98} questioned the analysis providing the 
condensation energy, and provided a different analysis
where a Gaussion fluctuation contribution was subtracted from the
experimental data. In Fig. \ref{fig:C(T)} the fit obtained by CKA to 
the specific heat of Tl2201 to 2D Gaussian fluctuation plus 
non-singular parts is reproduced. We will indicate
with an asterisk thermodynamic quantities from which the fluctuations 
have been subtracted. 
CKA calculated both results with ($E_n(0)-E^*(0)= 
25$ $\mbox{mJ/g-at.}$) and without ($E_n(0)-E(0) = 825$ $\mbox{mJ/g-at.}$) 
subtracting the singular part. The latter corresponds to the
earlier estimates in Refs. \onlinecite{pwa95,loram94,leggett96}. 
Following the basic idea of the ILT theory, CKA now equate 
the 'subtracted' value of the condenstaion energy to the decrease of the 
$c$-axis kinetic energy in the superconducting state, $\delta K$. Finally they 
use the standard relation 
\cite{marel96,leggett96,schuetzmann97,moler98,pwa98,tsvetkov98,kirtley98,basov99,basov01} 
between $\delta K$ and the $c$-axis penetration 
depth $\lambda_c$,  
\begin{equation} 
 {c^2 \over 8\lambda_c^2} \approx {\pi e^2 d^2\over 2 Ad  \hbar^2} \delta K, 
 \label{energy} 
\end{equation} 
Their resulting estimate of $\lambda_c$ is much larger
than the value $\lambda_c\simeq 1\mu$m derived previously in the context
of ILT. Vice versa the implication was, that the condensation energy
is approximately 40 times smaller than earlier estimates, which
had not corrected the specific heat for the singular part. 

The main source of this huge difference is the difficulty in determining
the 'normal' thermodynamic quantities.
The experimental questions include not only how 
but also {\em whether} the 'normal' state can be reached as a 
function of temperature, or indeed anything else. In other words, 
whether or not a parameter exists that can be tuned to lower the 
energy of the normal state below that of the equilibrium 
superconducting state so that it can be accessed without losing 
its fundamental character. Quite generally such a parameter does 
not always have to exist. For a weak coupling superconductor a 
magnetic field would suffice. In most cuprate superconductors the
required magnetic fields are beyond experimental reach, but Zn ions 
substituted for planar Cu may serve as an alternative for suppressing 
superconductivity\cite{loram1990,zn}. However, for the cuprates there is reason 
to believe that several 'normal' states are competing with the 
superconducting one (eg stripe, DDW, normal). In this case the 
field (or Zn-doping) required to mute superconductivity could be enough to 
rearrange the order between these 'normal' states, thus revealing 
the 'wrong' one when superconductivity gets suppressed. Thus we are confronted 
with the difficult situation, that the normal state 
entropy is not an experimental quantity, and can only be
determined based on theoretical considerations and/or by
extrapolating the normal state dependence as was indeed done
in Ref.~\onlinecite{cka98}, providing us, as we have seen, with
estimates of condensation energy differing by a factor of 40.

\begin{figure} 
        \centerline{\includegraphics[width=6cm,clip=true]{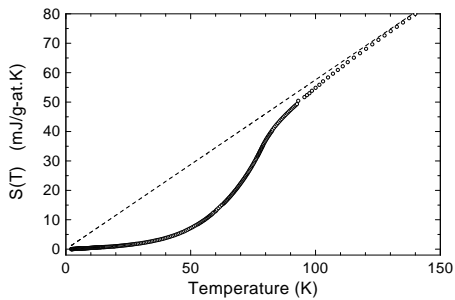}} 
        \caption{Experimental entropy versus temperature. 
        Dashed curve: Normal entropy with $\gamma=0.576$ mJ/G-at.K$^2$} 
\label{fig:S(T)} 
\end{figure} 
However, the situation isn't as bad as it looks. We
can use our knowledge of thermodynamics to
constrain the possible behaviour of $S_n(T)$:
Both $S(T)$ and $S_n(T)$ are
subject to the 2d and 3d law of thermodynamics. We can also
use the reasonable assumption, that for temperatures high enough, all 
superconducting correlations cease to exist, causing $S_n$ and $S$
to become equal in that limit. The circumstance 
that $F(T)$ corresponds to the equilibrium state, implies 
that for any temperature $F_n(T)$ has to be larger than $F(T)$. 
The corresponding constraints on the entropy are, in the same order,
\begin{equation}
 \begin{array}{l} 
 dS_n/dT > 0   \\
 S_{n}(0) = S(0) = 0    \\
 S_n(\infty) = S(\infty)    \\
 \int_0^{T}S_n(T')dT' \le \int_0^{T}S(T')dT' + E_{cond}
 \end{array}
 \label{eq:constraints}
\end{equation}
In Fig. \ref{fig:S(T)} the entropy is plotted as a function of
temperature. The condensation energy in this plot corresponds to the area 
between the equilibrium entropy ($S(T)$: open circles) and the 
normal entropy ($S_n(T)$: dashed curve). A condensation energy which is 
40 times smaller than the earlier estimates of the condensation energy\cite{pwa95,loram94,leggett96}, would require that the
area enclosed by $S_n(T)$ and $S(T)$ is also 40 times smaller than
in Fig. \ref{fig:S(T)}. Due the constraints of Eq. \ref{eq:constraints}, 
$S_n(T)$ would almost coincide with $S(T)$. $S_n(T)$ then 
has a sharp bend at $T_c$, which would 
correspond to a pathological phase transition of the normal state 
in close proximity to the real superconducting phase 
transition.

\begin{figure} 
        \centerline{\includegraphics[width=6cm,clip=true]{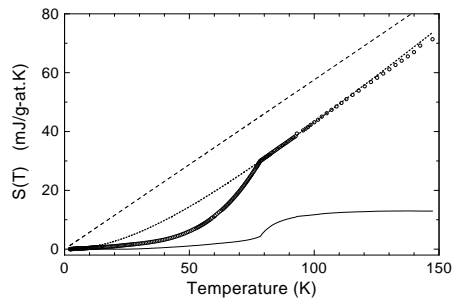}} 
        \caption{      
        Open circles: Corrected entropy versus temperature $S^*(T)=S(T)-S_{sing}(T)$, 
        following the procedure of Ref.~\onlinecite{cka98} where the singular 
        contribution was subtracted.
        Dashed curve: Normal entropy, $S_n(T)$, same as in Fig.\ref{fig:S(T)}. 
        Dotted curve: 'Normal' entropy, $S^*_n(T)$ fitted
        to the $T > T_c$-region of the corrected entropy curve. 
        Solid curve: Singular contribution to the entropy $S_{sing}(T)$.}
\label{fig:Scka(T)} 
\end{figure}

Let us now return to the analysis by CKA. If we calculate the 
entropy by integrating $C^*(T)/T$ with the singular part 
subtracted, we obtain the curve indicated with
open circles in Fig. \ref{fig:Scka(T)}. Above
the phase transition $S^*(T)$ approaches a linear 
temperature dependence with a negative offset, given by 
$S^*(T)=\gamma T-S_0$. We notice, that the corrected
entropy $S^*(T)$ and the normal entropy do not merge above $T_c$, causing 
the integral $E^*_{cond}=\int_0^{\infty}\left(S_n(T)-S^*(T)\right)dT$
to diverge.
We see, that subtracting the singular part has 
the effect that two procedures by which the condensation energy 
can be calculated provide opposite results:
  
$E_{cond}=\int_0^{\infty}\left(C(T)-C_n(T)\right)dT=25$$\mbox{mJ/g-at.}$  

$E_{cond}=\int_0^{\infty}\left(S_n(T)-S(T)\right)dT\rightarrow \infty$ 

This inconsistency does not arise in the earlier estimates of the condensation 
energy\cite{pwa95,loram94,leggett96}:
The fact that $S_n(T)$ and $S(T)$ merge above $T_c$ in Fig. \ref{fig:S(T)} 
guarantees that the same value of $E_{cond}$ is obtained with the two different 
formulas.

Following a recent suggestion by Abrahams\cite{elihuprivate} 
the CKA-analysis can be improved by adopting a specific heat 
exponent $\eta$ different from 1 below $T_c$. We must also take
into account that the entropy above for $T>T_c$ approaches 
$S^*(T)=\gamma T-S_0$. A simple analytical expression
with these two properties is
\begin{equation}
  S^*_n(T)= S_0
  \left\{\left(1+\left(\frac{T}{T_0}\right)^{\eta}\right)^{1/\eta}-1\right\}
\label{eq:s*n}
\end{equation}
We have tried to fit different values for $\eta$. If we choose 
$\eta \ge 2.5$ the 'normal' entropy $S^*_n(T)$ becomes 
smaller than the experimentally measured $S(T)$ for
temperatures below 40 K. Although such a possibility can not be excluded by the 
thermodynamical constraints of Eq. \ref{eq:constraints},  
from a microscopic perspective it looks suspicious that
the entropy of the (gapped) superconducting state could exceed that of the
normal state. To avoid this, we have adopted the value $\eta = 2.0$. The 
best fit in the region between $T_c$ and $110 K$ was obtained with 
$S_0=34.85$ $\mbox{mJ/g-at.}$K and $T_0 = 50 K$.
In Fig.\ref{fig:Scka(T)} we display the singular contribution to the
entropy $S_{sing}(T)$,
the corrected entropy $S^*(T)=S(T)-S_{sing}(T)$, and the normal
state entropy $S^*_n(T)$. Note that if indeed it would be justified to
subtract the singular contribution, the presence of a negative offset
in the entropy implies that this 'normal' state would have a 
pseudo-gap\cite{loram94,loram2001}.     

\begin{figure} 
        \centerline{\includegraphics[width=6cm,clip=true]{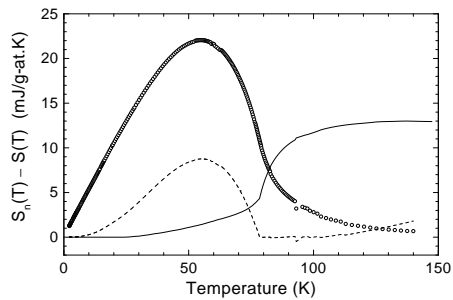}} 
        \caption{Open circles: Normal state entropy minus experimental entropy.
        Solid curve: Singular contribution to the entropy. 
        Dashed curve: 'Normal' entropy $S^*_n(T)$, minus
        the corrected entropy $S^*(T)$.} 
\label{fig:Sdif(T)} 
\end{figure}
\begin{figure} 
        \centerline{\includegraphics[width=6cm,clip=true]{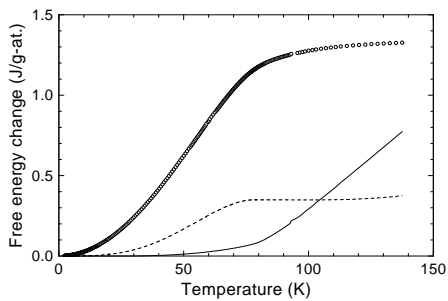}}
        \caption{Integrated entropy differences of Fig. \ref{fig:Sdif(T)}.
        Open circles: $\int_0^T (S_n(T')-S(T'))dT'$.
        Dashed curve: $\int_0^T (S^*_n(T')-S^*(T'))dT'$.
        Solid curve: Singular contribution to the free energy. 
        } 
\label{fig:F(T)} 
\end{figure}

 
In Fig. \ref{fig:Sdif(T)} we display both $S_n(T)-S(T)$, and the quantity
which corresponds to the improved version of the CKA-analysis, $S^*_n(T)-S^*(T)$. 
We see, that with the improved CKA-analysis the difference entropy 
between the normal and superconducting state indeed becomes zero above $T_c$. We
carried the analysis of the specific heat one step further, and calculated the
free energy difference curves. These are displayed in Fig. \ref{fig:F(T)}. The
corrected CKA-analysis thus provides a condensation energy
of $0.35$ \mbox{J/g-at}. This value is about 30 percent of the
condensation energy from direct integration of the total entropy of Fig.\ref{fig:S(T)}
which gives $E_{cond} = 1.3$ \mbox{J/g-at.}.  We see, that even if it would be justified
to subtract the fluctuations,  the correct analysis would still reduce the
condensation by a factor of 2 or 3, not by a factor of 40 as stated in Ref.~\onlinecite{cka98}.

However, removing the singular contribution from the experimental data comes 
with a penalty: As a reminder that a fluctuation contribution
has been subtracted in the CKA-analysis, we have displayed in Figs. \ref{fig:Scka(T)},
\ref{fig:Sdif(T)} and \ref{fig:F(T)} its contribution to the entropy and the
free energy. We see, that the fluctuation entropy has a conspicious step at 
$T_c$, which indicates that these fluctuations are intimately 
connected to the superconducting phase transition. The fact that the fluctuation 
specific heat is strongly peaked at the phase transition, 
implies that there is an additional reduction of the internal energy 
due to the fluctuation contribution. Hence it 
would seem to be overwhelmingly natural to count this part when
one estimates the energy by which the superconducting state
is stabilized relative to competing (non-superconducting) phases.

In conclusion, we have shown, that in the original analysis of CKA, {\em i.e.} 
subtraction of a fluctuation contribution to the specific heat, 
a corrected entropy below Tc was used which does not match 
the value used above $T_c$. This internal inconsistency was the main 
reason why the condensation energy was estimated a 
factor of 40 smaller than the value obtained in earlier 
publications. This problem could have been fixed by letting 
the specific heat exponent be less than one at low temperatures. We
have repeated the CKA-analysis with this fix, resulting
in a condensation energy which is ten times as large.
However, it seems overwhelmingly natural to include the contribution of
the fluctuations in the analysis, because experimentally the fluctuation 
contribution to the internal energy appears to be intimately linked to the 
superconducting phase transition. This results in a condensation energy of 
approximately $1.3$ \mbox{J/g-at.} for optimally doped Tl2201. 

We gratefully acknowledge stimulating discussions with K. A. Moler, S. Kivelson, J. Zaanen,
S. Chakravarty, H.-Y. Kee, and E. Abrahams during the preparation of this paper.
\end{document}